\documentclass[10pt]{iopart}
\usepackage{graphicx}

\begin{document}

\title{Gutzwiller variational theory for the Hubbard model
with attractive interaction}

\author{J\"org B\"unemann\dag, Florian Gebhard\dag ,
 Katalin Radn\'oczi\ddag, and Patrik Fazekas\ddag}
\address{\dag\ Fachbereich Physik, Philipps-Universit\"at Marburg,
D--35032 Marburg, Germany \\
\ddag\ Research Institute for Solid-State Physics and Optics,
Hungarian Academy of Sciences, H-1525 Budapest, Hungary}


\begin{abstract}%
We investigate the electronic and superconducting properties 
of a nega\-tive-$U$ Hubbard model. For this purpose we evaluate 
a recently introduced variational theory based on Gutzwiller-correlated
BCS wave functions.
We find significant differences between our approach and standard
BCS theory, especially for the superconducting gap. 
For small values of $|U|$, we derive analytical ex\-press\-ions for the 
order parameter and the superconducting gap which we compare 
to exact results from perturbation theory.   
\end{abstract}

\submitto{\JPCM}


\section{Introduction}

Gutzwiller-correlated wave functions~\cite{Gutwiller} 
were originally proposed for 
a variational examination of the one-band Hubbard model~\cite{Hubbard}
\begin{equation}\label{Hamiltonian}
\hat{H}=\sum_{i\neq j}\sum_{\sigma=\uparrow,\downarrow} 
t_{i,j}\hat{c}_{i\sigma}^{+}\hat{c}_{j\sigma}^{}+
U\sum_{i}\hat{n}_{i\uparrow}\hat{n}_{i\downarrow}=\hat{H}_{\rm kin}+\hat{H}_{U}
\end{equation}
with positive Coulomb repulsion~$U$.
Here, $i$ and~$j$ are sites of the respective lattice under investigation
and $\hat{c}_{i\sigma}^{+}$ ($\hat{c}_{i\sigma}$) 
creates (annihilates) an electron with spin $\sigma=\uparrow,\downarrow$.
As usual, $\hat{n}_{i\sigma}=\hat{c}_{i\sigma}^{+} \hat{c}_{i\sigma}$
counts the number of $\sigma$-electrons on site~$i$.

Gutzwiller's primary intention was to study ferromagnetism in 
such a system. Unfortunately, the evaluation of expectation values 
for Gutzwiller's variational 
wave function still poses a generally unsolvable many-particle problem. 
Gutzwiller introduced an approximate evaluation scheme, 
the so called {\sl Gutzwiller approximation}, which was based on
phenomenological counting arguments; for a mathematically sound 
introduction to this technique, see~\cite{JBcounting}. 
More recent research on 
Gutzwiller wave functions focused on the development 
of better controlled evaluation schemes.
Main progress was made in the limit of 
infinite spatial dimensions~\cite{Gebhard1,Gebhard2}
where the Gutzwiller approximation
turned out to be exact in some cases~\cite{MetznerVollhardt}.

Despite its general importance as the most 
simplistic model for correlated electron systems, 
the one-band Hubbard model is only of limited use
for the investigation of real materials. Therefore, a second aim was  
the evaluation of more general Gutzwiller wave functions for multi-band
models. Such a theory has first been derived in~\cite{PRB98} and 
was later used in numerical studies on Nickel~\cite{EPL,springer}.
The results of these nu\-meri\-cal investigations 
are in excellent agreement with experiments and pose a major improvement 
over band-structure calculations based on the local-spin-density 
ap\-proxi\-mation
to density-functional theory.

Since the advent of high-temperature superconductivity about 20~years ago,
Gutzwiller wave functions have been used for the 
investigation of systems with alleged attractive interactions, 
most notably the two-dimensional $t$-$J$-model~\cite{zhang1,zhang2,anderson}. 
For the evaluation of those Gutzwiller-correlated BCS wave 
functions approximations have been used which were
based on counting arguments similar to the original Gutzwiller approximation. 
In~\cite{springer} we developed a general theory for the evaluation of 
superconducting multi-band Gutzwiller wave functions in infinite spatial
dimensions. An application of these general results to real materials 
requires  significant numerical efforts.
In this report we present first results 
on the simplest correlated electron system that exhibits superconductivity, 
the attractive Hubbard model. The same system has been studied  
using Gutzwiller-correlated wave functions in~\cite{suzuki},
based on an evaluation scheme first derived in~\cite{metzner}. 
For the attractive one-band Hubbard model with a Gaussian density of states, 
the numerical results of both methods seem to agree. Unfortunately,
we cannot provide a rigorous proof for the equivalence 
of both approaches. Since the evaluation scheme used in~\cite{suzuki,metzner} 
cannot be generalized easily to more realistic model systems,
we do not further pursue their approach. 

We structure our article as follows. In section~\ref{chap2} we introduce 
the Gutzwiller-correlated superconducting wave functions 
for the negative-$U$ Hubbard model
and derive analytical expressions for the variational ground-state 
energy. The minimisation of this energy is discussed in section~\ref{chap3}. 
In particular, we derive an effective one-particle Hamiltonian
which provides a link to ARPES experiments. In section~\ref{chap4}
we focus on the half-filled Hubbard model. We discuss the small-$|U|$ limit
analytically and provide numerical results for a system with a 
semi-elliptic density of states. A summary in section~\ref{chap5} closes our
presentation.

\section{Superconducting Gutzwiller wave functions}\label{chap2}
We consider the Gutzwiller wave function
\begin{equation}\label{wavefunc}
| \Psi_{\rm G}\rangle \equiv \prod_{i}  \hat{P}_{i} | \Psi_0\rangle \; .
\end{equation}
Here, $| \Psi_0\rangle$ is a normalized
quasi-particle vacuum and the local correlator
\begin{equation}
\hat{P}_{i}\equiv\hat{P}=\sum_{I,I'}\lambda_{I,I'}| I\rangle_{i}
{}_{i}\langle I'|
\end{equation}
induces transitions at the lattice site $i$ between the four atomic states 
$| I\rangle$, i.e., the empty state $|\emptyset\rangle$, 
the doubly occupied state $|d\rangle$ and the two single electron states 
$|\sigma\rangle=|\uparrow\rangle,|\downarrow\rangle$. We denote expectation 
values with respect to $| \Psi_{\rm G}\rangle$ and $| \Psi_0\rangle$ as 
$\langle \cdots \rangle_{\rm G}$ and $\langle \cdots \rangle_{0}$, 
respectively. 
In general, there are
16~variational parameters $\lambda_{I,I'}$. When we assume that 
there is no magnetic or charge order in the system, it is sufficient to 
work with only four independent and real parameters: 
(i)~$\lambda_{\emptyset}\equiv\lambda_{\emptyset,\emptyset}$, 
(ii)~$\lambda_{1}\equiv\lambda_{\uparrow,\uparrow}
=\lambda_{\downarrow,\downarrow}$,
(iii)~$\lambda_{d}\equiv\lambda_{d,d}$, 
(iv)~$\lambda_{B}\equiv\lambda_{d,\emptyset}=\lambda_{\emptyset,d}$. 
It can be shown that a more general Ansatz  with complex parameters 
$\lambda_{I,I'}$ does not lead to a variational improvement.  
As shown in \cite{springer} the four parameters are not independent.
Instead, they obey the constraints
\begin{eqnarray}\nonumber
1&=&\langle\hat{P}_{i}^2 \rangle_0  \\ \label{cond2}
n_{0}&=&\langle \hat{c}_{i\sigma}^{+}\hat{P}^2
\hat{c}_{i\sigma}^{}  \rangle_0\;, \\ \nonumber
\Delta_{0}&=&\langle \hat{c}_{i\uparrow}^{+}
\hat{P}^2\hat{c}_{i\downarrow}^{+}  \rangle_0\;.
\end{eqnarray}
Here, $n_{0}$ and $\Delta_{0}$ are the elements of the 
uncorrelated local 
density matrix
\begin{equation}\label{densmat}
n_{0}= \langle  \hat{c}_{i\sigma}^{+}  \hat{c}_{i\sigma}^{}\rangle_0
\;\;,\;\;\Delta_{0}= 
\langle  \hat{c}_{i\uparrow}^{+}  \hat{c}_{i\downarrow}^{+}\rangle_0
\end{equation}
which we  assume to be independent of the site index 
$i$ and the spin index $\sigma$.  
The evaluation of equations~(\ref{cond2}) leads to
 \begin{eqnarray} \nonumber
1&=&(\lambda_{\emptyset}^2+\lambda_{B}^2)m_{\emptyset,0}
+2\lambda_{1}^2m_{1,0}
+(\lambda_{\emptyset}^2+\lambda_{B}^2)d_{0}
+2\Delta_{0}(\lambda_{\emptyset}+
\lambda_{d})\lambda_{B}\;,\\ 
n_{0}&=&(\lambda_{\emptyset}^2+\lambda_{B}^2)m_{1,0}+\lambda_{1}^2d_0\;,\\
 \nonumber
\Delta_{0}&=&\lambda_{1}^2\Delta_{0}-(\lambda_{\emptyset}+\lambda_{d})
\lambda_{B}m_{1,0}\; ,
\end{eqnarray}
with the expectation values 
 \begin{eqnarray}\nonumber
d_0&\equiv& \langle \hat{n}_{i\uparrow}\hat{n}_{i\downarrow} 
 \rangle_0=n_0^2+\Delta_0^2\;,\\
m_{1,0}&\equiv& \langle \hat{n}_{i\uparrow}(1-\hat{n}_{i\downarrow}) 
\rangle_0=n_0-d_0\;,\\\nonumber
m_{\emptyset,0}&\equiv&\langle (1-\hat{n}_{i\uparrow})
(1-\hat{n}_{i\downarrow}) \rangle_0=1-2m_{1,0}-d_0\;.
\end{eqnarray}
We use these equations in order to express the three
parameters $\lambda_{\emptyset} $, $\lambda_{1} $, $\lambda_{B} $  
 in terms of the fourth parameter $\lambda_{d}$, 
\begin{eqnarray}\nonumber
\lambda_1^2&=&1+x(d_0-n_0)\;,\\
\lambda_{\emptyset}^2&=&\lambda_d^2-x(1-2n_0)\;,\\ \nonumber
\lambda_B^2&=&1-\lambda_d^2+x m_{\emptyset,0}\;.
\end{eqnarray} 
Here, we introduced the abbreviation
\begin{equation}
x=\frac{B}{2A^2}\left(\sqrt{1-\frac{4A^2C}{B^2}}
-1  \right)
\end{equation}
with
\begin{eqnarray}\nonumber
A&=&\Delta_0^2+(1-2n_0)m_{\emptyset,0}\;,\\
B&=&-4\lambda_d^2\Delta_0^2m_{\emptyset,0}+2A(1-\lambda_d^2)(1-2n_0)\;,\\
\nonumber
C&=&(1-\lambda_d^2)\left((1-2n_0)^2(1-\lambda_d^2)-4\Delta_0^2\lambda_d^2\right)\;.
\end{eqnarray} 

The expectation value of the Hamiltonian~(\ref{Hamiltonian}) 
for the Gutzwiller wave function~(\ref{wavefunc}) can be 
evaluated in the limit of infinite spatial dimensions \cite{springer}. 
For a hopping term this evaluation yields
\begin{eqnarray}
\langle \hat{c}_{i,\sigma}^{+} \hat{c}_{j,\sigma}^{} \rangle_{\rm G}&=&
q^2\langle \hat{c}_{i,\sigma}^{+} \hat{c}_{j,\sigma}^{} \rangle_{0}
+\bar{q}^2\langle \hat{c}_{i,\sigma}^{} \hat{c}_{j,\sigma}^{+} \rangle_{0}
\nonumber \\
&&
+q\bar{q}\left(\langle \hat{c}_{i,\sigma}^{+}\hat{c}_{j,\sigma}^{+}\rangle_{0}
+\langle \hat{c}_{i,\sigma}^{}\hat{c}_{j,\sigma}^{}\rangle_{0}\right)
\end{eqnarray}
where
\begin{eqnarray}
q&=&q(\lambda_{d},n_0,\Delta_0)=\lambda_{1}(\lambda_{\emptyset}(1-n_0)+\lambda_{d}n_0)
+2\lambda_{B}\Delta_0\lambda_{1}\;,\\
\bar{q}&=&\bar{q}(\lambda_{d},n_0,\Delta_0)=\Delta_0\lambda_{1}(\lambda_{d}
-\lambda_{\emptyset}
+\lambda_{B})(1-2n_0)\;
 \end{eqnarray}
are renormalization factors  for normal and anomalous hopping.
The expectation value of the Hamiltonian~(\ref{Hamiltonian}) then reads
\begin{equation}
E^{\rm var}=\langle  \hat{H} \rangle_{\rm G}=
Q\langle\Psi_0| \hat{H}_{\rm kin}|\Psi_0 \rangle+Ud
\equiv Q E^0_{\rm kin}+U d \;,
\end{equation}
with
\begin{equation}
Q\equiv q^2-\bar{q}^2\quad ,\quad 
d=\lambda_{d}^2d_0+\lambda_{B}^2m_{\emptyset,0}
+2\lambda_{B}\lambda_{d}\Delta_0 \; .
\end{equation}

\section{Minimization procedure}\label{chap3}
The variational energy has to be minimized with respect to 
$\lambda_{d}$ and $| \Psi_0\rangle $ whereby the equations~(\ref{densmat})
and the normalization of  $| \Psi_0\rangle $ 
need to be obeyed. For this purpose  we introduce Lagrange parameters 
$\eta_{{\rm n}}$, $ \eta_{{\rm s}}$ and $E^{\rm SP}$. 
Furthermore, we  keep the 
average number of particles
\begin{equation}
\bar{n}=\frac{N}{L}\equiv n(\lambda_{d},n_0,\Delta_0)
= (2\lambda_{1}^2m_{1,0}+2d)
\end{equation}
fixed by means of a Lagrange parameter $\mu$. 
The minimisation problem then becomes
 \begin{eqnarray}\nonumber
E_0^{\rm var}&=&
\mathop{\rm Minimum}_{\lambda_{d},\eta_{{\rm n}}, \eta_{{\rm s}},
\mu,|\Psi_0\rangle}
\Bigl[E^{\rm var}
-E^{\rm SP}\left(\langle \Psi_0 |\Psi_0\rangle -1\right)\\
&&-\eta_{\rm s}\sum_{i,\sigma}
(\Delta_0-\langle\Psi_0| \hat{c}^{+}_{i,\uparrow} \hat{c}^{+}_{i,\downarrow} 
|\Psi_0 \rangle+{\rm c.c.})\\
&&-\eta_{\rm n}\sum_{i,\sigma}
(n_0-\langle\Psi_0| \hat{c}^{+}_{i,\sigma} \hat{c}^{}_{i,\sigma} 
|\Psi_0 \rangle)+\mu L(\bar{n}-n(\lambda_{d},n_0,\Delta_0))\Bigr]\; .
\nonumber
\end{eqnarray}
 The minimization with respect to $|\Psi_0\rangle$ 
can be carried out explicitly 
and leads to an effective Schr\"odinger equation in momentum space,
 \begin{eqnarray}\label{seq}
\hat{H}^{\rm eff}|\Psi_0\rangle&=&E^{\rm SP}|\Psi_0\rangle\; ,\\
\hat{H}^{\rm eff}&=&\sum_{k,\sigma}\epsilon_{k}
\hat{c}^{+}_{k\sigma}\hat{c}^{}_{k\sigma}
+\eta_{\rm s}\sum_{k}(\hat{c}^{+}_{k\uparrow}
\hat{c}^{+}_{-k\downarrow}+{\rm h.c.})\; ,
\end{eqnarray}
with 
\begin{eqnarray}\nonumber
\epsilon_{k}&=&\eta_{\rm n}+Q\epsilon_{k}^0 \;,\\
\epsilon_{k}^0&=&\frac{1}{L}\sum_{i,j}t_{i,j}e^{-{\rm i}k(i-j)}\;.
\end{eqnarray}
The one-particle Schr\"odinger equation~(\ref{seq}) is readily solved,
\begin{equation}\label{seq2}
\hat{H}^{\rm eff}=\sum_{k}E_{k}\left(\hat{h}^{+}_{k,0}\hat{h}^{}_{k,0}
+\hat{h}^{+}_{k,1}\hat{h}^{}_{k,1}\right)+{\rm const.}\;,
\end{equation}
by means of a Bogoliubov transformation 
\begin{eqnarray}\nonumber
\hat{c}_{k,\uparrow}=u_{k}\hat{h}_{k,0}+v_{k}\hat{h}^{+}_{-k,1}\;,  \\
\hat{c}_{-k,\downarrow}^{+}=-v_{k}\hat{h}_{k,0}+u_{k}\hat{h}^{+}_{-k,1}\;.
\end{eqnarray}
Here, the real coefficients 
$u_{k}$, $v_{k}$ and the energies $E_{k}$ are determined by the equations
\begin{equation}
E_{k}= \epsilon_{k}(u_{k}^2-v_{k}^2)-2\eta_{\rm s}u_{k}v_{k} \;,
\end{equation}
and
\begin{equation}
2\epsilon_{k}u_{k}v_{k}+\eta_{\rm s}(u_{k}^2-v_{k}^2)=0 \quad, \quad
u_{k}^2+v_{k}^2 =1 \; ,
\end{equation}
which are solved by 
\begin{eqnarray}\label{ek}
E_{k}&=&{\rm sign}(\epsilon_{k})\sqrt{\epsilon_{k}^2+\eta_{\rm s}^2}
={\rm sign}(\epsilon_k^0)\sqrt{(Q\epsilon_k^0
+\eta_{\rm n})^2+\eta_{\rm s}^2}
\;,\\\nonumber
u_{k}&=&\frac{1}{\sqrt{2}}\sqrt{1+\frac{\epsilon_{k}}{E_{k}}}\;,\\\nonumber
v_{k}&=&-{\rm sign}(\epsilon_{k} \eta_{\rm s})\frac{1}{\sqrt{2}}
\sqrt{1-\frac{\epsilon_{k}}{E_{k}}}\;.
\end{eqnarray}
The one-particle state $|\Psi_0\rangle$ is chosen as the 
ground state of~(\ref{seq2})
\begin{equation}
|\Psi_0\rangle=\prod_{k(E_{k}<0)}h_{k,0}^{+}h_{k,1}^{+}|{\rm vac}\rangle\;.
\end{equation}
In \cite{springer,thul} it was shown that the eigenvalues~(\ref{ek})
in the effective Hamiltonian~(\ref{seq2})
 can be interpreted as  quasi-particle energies that might be measured in  
ARPES experiments. Therefore,  the parameter  $\eta_{\rm s}$ describes the 
superconducting gap and  $Q$ is a measure for the band-width 
renormalization. 

When we introduce the uncorrelated density of states 
\begin{equation}
D{(\tilde{\epsilon})}=\frac{1}{L}\sum_{k}\delta(\tilde{\epsilon}-\epsilon_{k})
\end{equation}
the elements of the local density matrix~(\ref{densmat}) can be written as 
\begin{eqnarray}\label{con1}
n_0&=&\frac{1}{2}\int d\tilde{\epsilon}D(\tilde{\epsilon})
\left(1-\frac
{Q\tilde{\epsilon}+\eta_{\rm n}}
{\sqrt{(Q\tilde{\epsilon}+\eta_{\rm n})^2+\eta_{\rm s}^2}          }
\right)\;,\\\label{con2}
\Delta_0&=&\frac{\eta_{\rm s}}{2}\int d\tilde{\epsilon}
D(\tilde{\epsilon})\frac{1}
{   \sqrt{(Q\tilde{\epsilon}+\eta_{\rm n})^2+\eta_{\rm s}^2}        }\;,
\end{eqnarray}
and the one-particle energy reads 
\begin{equation}
E_{\rm kin}=
Q\int d\tilde{\epsilon} 
D(\tilde{\epsilon})\tilde{\epsilon}
\left(1-\frac{Q\tilde{\epsilon}+\eta_{\rm n}}
{\sqrt{(Q\tilde{\epsilon}+\eta_{\rm n})^2
+\eta_s^2}}\right)
= Q E_{\rm kin}^0\; .
\label{ekinint}
\end{equation}

\section{Results for half band-filling}\label{chap4}
We assume that the density of states is symmetric and focus on the 
half-filled case 
in which certain simplifications occur. 
By setting $\eta_{\rm n}=0$ we ensure that  $2n_0=n=1$ and 
$\Delta \equiv\langle 
c_{i\uparrow}^+c_{i\downarrow}^+\rangle_{\rm G}=\Delta_0$.
The remaining numerical task is the  minimisation of the variational energy
\begin{equation}
E_{\rm var}=Q(\Delta_0,\lambda_d)E_{\rm kin}^{0}(\Delta_0,\eta_s)
+U d(\Delta_0,\lambda_d)
\end{equation} 
with respect to $\eta_s$ and $\lambda_d$  where $\Delta_0$ is given by
equation~(\ref{con2}). In a pure mean-field BCS theory 
we have to set $\lambda_d=1$ 
and minimise the energy  only with respect to $\eta_s$. Note that for the 
numerical minimisation we found it more convenient to work with the two 
variational parameters $d$ and $y\equiv \eta_{\rm s}/(2Q)$.

\subsection{Comparison with perturbation theory}
For small values of $U$, and, correspondingly, small values 
of $\eta_{\rm s}$ and $\Delta_0$, the minimisation can be carried out
analytically. 
To leading order in~$\eta_{\rm s}$ and~$U$ the variational energy 
reads
\begin{eqnarray}
E_{\rm var}(\eta_{\rm s})
&\approx&\epsilon_0-\frac{|U|}{4}-D(0)\eta_{\rm s}^2
\ln{    \left(\frac{2\eta_{\rm s}}{W}\right)        }+
U\alpha_U D(0)^2\eta_{\rm s}^2
\left[\ln{\left(\frac{2\eta_{\rm s}}{W}\right)}\right]^2 \nonumber \\
&& \label{ensmallu}
\end{eqnarray}
with the band-width $W$ and the bare kinetic energy 
\begin{equation}
\epsilon_0=2\int_{-\infty}^{0}\epsilon D(\epsilon) d\epsilon \;.
\end{equation}
The difference between a Gutzwiller and a BCS wave function shows up 
solely in the respective values of the coefficient 
$\alpha_U$ in~(\ref{ensmallu}). It is
$\alpha_U=1$ in standard BCS theory and $\alpha_U=1-(3U)/(16\epsilon_0)<1$ 
for a Gutzwiller-correlated BCS state. The minimisation with respect to 
 $\eta_{\rm s}$ then yields 
\begin{equation}
\eta_{\rm s}=\frac{W}{2}
\exp{\left(-\frac{1}{2}-\frac{1}{D(0)|U|\alpha_{U}}\right)}\;.
\end{equation}
In the limit $U\rightarrow 0$
the order parameter $\Delta\sim \eta_{\rm s}$ 
in the Gutzwiller theory is therefore 
just renormalized in comparison with the respective BCS parameter
\begin{equation}
 \Delta^{\rm GW}/ \Delta^{\rm BCS}\equiv r
=\exp{\left(-\frac{C}{D(0)}\right)}
\;. \label{renorm}
\end{equation}
An exact calculation of $\Delta$ by means of perturbation 
theory~\cite{peter,comment} 
yields the same functional form, eq.~(\ref{renorm}), however, with a different
 parameter $C$. It is  
$C=3/(16|\epsilon_0|)$
in the Gutzwiller theory and
\begin{equation}\label{cexact}
C^{\rm exact}=2\int_{0}^{\infty}\int_{0}^{\infty}d\epsilon d\epsilon'
\frac{D(\epsilon)D(\epsilon')}{\epsilon+\epsilon'}
\end{equation}
in the exact evaluation. 
In the rest of our work we employ a semi-elliptic density of states 
of width $W=4t$,
\begin{equation}
D(\epsilon)=\frac{1}{2\pi}\sqrt{4-\epsilon^2} \; ,
\end{equation}
which is realized in a Bethe lattice with infinite coordination number.
For this semi-elliptic density of states
we find $C_{\rm semi}=9\pi /128\approx 0.221$ and  
$C_{\rm semi}^{\rm exact}=4/(3\pi)\approx 0.424$.
We then have
\begin{equation}
r_{\rm semi}=\exp{\left(-\frac{9 \pi^2}{128}\right)}\approx 0.4996\;,
\end{equation}
in comparison with the exact value 
$r_{\rm semi}^{\rm exact}\approx 0.2636$.

Another quantity of interest is the condensation energy, i.e.,
the energy difference between the superconducting and the 
normal ground state. Both in our approach and in perturbation theory,  
one finds to leading order in~$U$ 
\begin{equation}
 E_{\rm cond}\approx -\frac{1}{2}D(0)U^2\Delta^2=r^2 E^{\rm HF}_{\rm cond}
\; .
\end{equation}
For the semi-elliptic density of states the condensation energy is 
therefore over\-es\-ti\-mated by a factor of $(r/r^{\rm exact})^2$ which
shows that BCS theory overestimates the condensation energy
by a factor of about 14 for a semi-elliptic density of states
whereas Gutzwiller theory is only off by a factor of four. 

In the large $U$ limit, 
the BCS energy shows an unphysical divergence of the 
condensation energy, $E_{\rm cond}^{\rm BCS}\sim U$. 
Our Gutzwiller approach yields the correct $1/U$ behaviour,
\begin{equation}
E_{\rm cond}=-\frac{\tilde{c}}{U}\;.
\end{equation}
However, the coefficient
\begin{equation}
\tilde{c}=2\int_{-\infty}^0D(\epsilon)\epsilon^2 d\epsilon
\end{equation}
is twice as large as the exact result. This deficit is due to the 
poor description of the paramagnetic insulator in the Gutzwiller theory. 

\begin{figure}[htb]
\centerline{\resizebox{13cm}{!}{\includegraphics{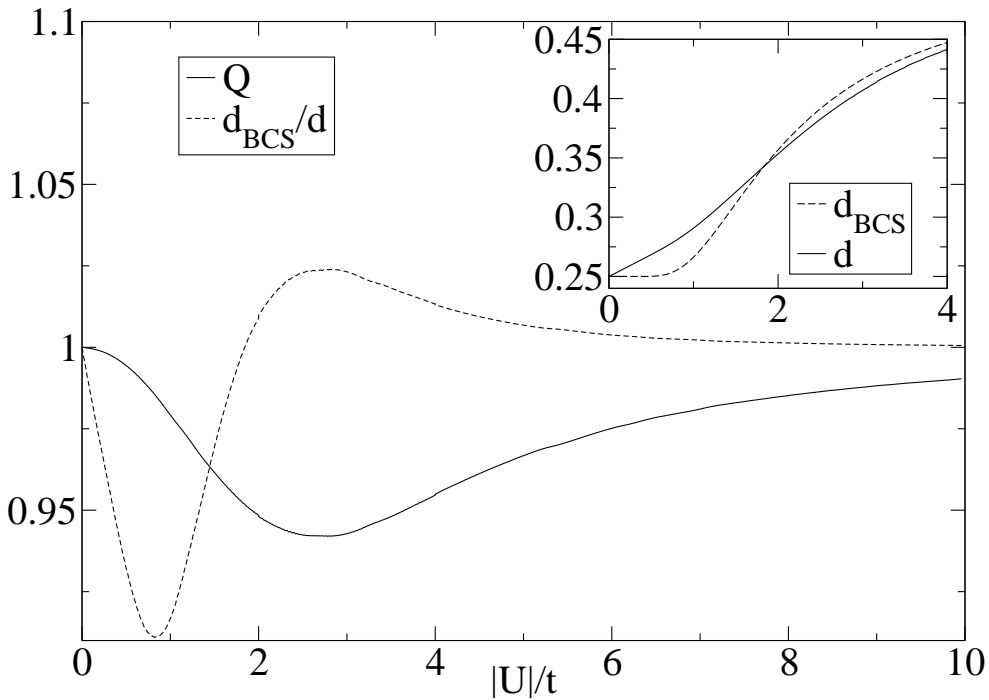}}}
\caption{Renormalisation factor~$Q$ (solid line) 
and ratio of double 
occupancies $d_{\rm BCS}/d$ (dashed line) as a function of $|U|/t$.
Inset: double occupancies~$d$ (solid line) and  $d_{\rm BCS}$ 
(dashed line).\label{fig1}}
\end{figure}

\subsection{Numerical results}

As seen in eq.~(\ref{ek}) the Gutzwiller--Bogoliubov quasi-particles have a
dispersion relation $|E_{k}|= \sqrt{(Q\epsilon_k^0)^2+\eta_{\rm s}^2)}$.
The band-width renormalization factor $Q$  in the Gutzwiller theory deviates 
only slightly from the 
uncorrelated BCS value $Q=1$, as shown in Fig.~\ref{fig1}. 
It has a minimum $q_{\rm min}\approx 0.95$ for medium-size 
correlation strength $|U|\approx 3$.  In the limit 
$U\rightarrow -\infty$ the BCS wave function itself has the optimum 
expectation value $d \approx  1/2$ of double occupancies such that
the Gutzwiller correlator $\hat{P}$ cannot achieve any  further energy gain. 
Therefore, the renormalization approaches unity in this limit.
In Fig.~\ref{fig1} we 
also plot the average double occupancy in Gutzwiller and BCS theory 
(see inset) as well as the ratio of these two quantities.
For small values of $|U|$ the double occupancies
show a qualitatively different  behaviour in both approaches. 
In Gutzwiller
theory we find the well-known linearity
$(d-1/4) \sim U$ whereas BCS theory gives the typical exponential dependence,
$(d_{\rm BCS}-1/4) \sim \Delta_0^2 \sim 
 \exp[-(1/|U|)]$. The BCS value $d_{\rm BCS}$
is smaller than $d$ only for small values of $|U|$. 
For larger $|U|$ the limited flexibility
of the BCS wave function leads to an overshooting of the double occupancy.

\begin{figure}[htb]
\centerline{\resizebox{13cm}{!}{\includegraphics{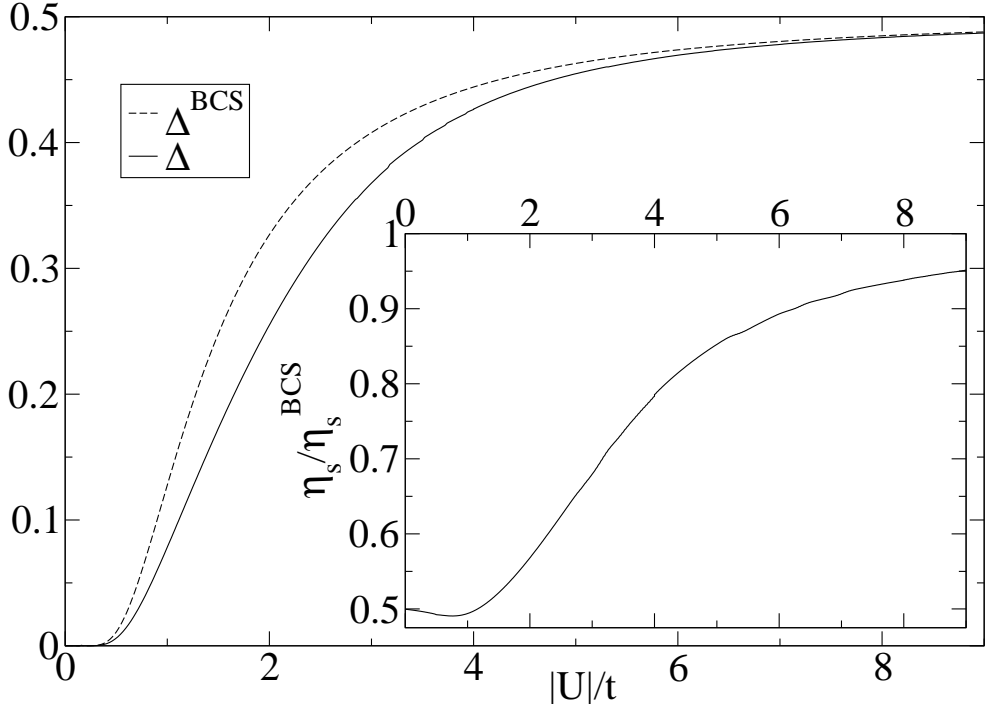}}}
\caption{Superconducting order parameters for the Gutzwiller wave function
 (solid line) and the BCS wave function (dashed line) as a 
function of $|U|/t$. 
Inset: Ratio of the 
superconducting gaps in the  Gutzwiller and the BCS theory.\label{fig2}}
\end{figure}

In Fig.~\ref{fig2}, we show the values of the superconducting order parameter 
$\Delta$ in both theories as a function of $U$. The inset of this figure  
shows the ratio of the two 
superconducting gaps. Note that in BCS theory we have
$\eta^{\rm BCS}_{\rm s}=|U| \Delta^{\rm BCS}$.  The order parameters
 are quite similar in both approaches apart from small values of $|U|$ where
\begin{equation}
 \lim_{|U|\rightarrow 0}\frac{\Delta}{\Delta^{\rm BCS}}=
 \lim_{|U|\rightarrow 0}\frac{\eta_{\rm s}}{ \eta_{\rm s}^{\rm BCS}}
=r  \; .
\end{equation}
Other than the order parameter, the superconducting gaps in both theories  
differ significantly over a large range of correlation parameters~$|U|$.  
Equation~(\ref{renorm}) shows that BCS theory fails 
when the density of states at the Fermi energy~$D(0)$ becomes small.

\section{Summary}\label{chap5}
As a first application of our recently developed Gutzwiller theory for 
superconducting systems~\cite{springer} 
we investigated the attractive Hubbard model. 
Our variational approach leads to the diagonalisation of an effective 
one-particle Hamiltonian, very similar to the corresponding mean-field 
BCS theory. Quantitatively, however, the resulting quasi-particle 
band structure differs remarkably from the BCS band
energies. This observation holds, in particular, for the superconducting gap
in both theories. 

In the limit of small Coulomb interaction~$|U|/t$, 
a comparison of our approach with perturbation theory shows
that Gutzwiller theory reproduces the renormalization 
of the superconducting order parameter qualitatively. 
Quantitatively, the results 
depend on the bare density of the states of the system. 
For the semi-elliptic density
of states, for example, our renormalization of the gap 
is too small by a factor of about two.

The condensation energy, i.e., the  energy difference between 
the paramagnetic and the superconducting 
ground state, is another quantity which is notoriously overestimated
in BCS theory. This holds, in particular, for large~$U$ where 
the condensation energy should approach zero but diverges in BCS theory.
In contrast, our Gutzwiller theory gives the correct order of 
magnitude for the condensation energy for all Coulomb strengths. 

Apparently, the investigation of superconducting 
systems with Gutzwiller wave functions leads to a significant 
improvement over on simple BCS-type calculations. Therefore, our approach 
should be a useful tool for the investigation of more realistic model systems 
that exhibit superconductivity.

\end{document}